# Localized waves supported by the rotating waveguide array

Xiao Zhang,[1] Fangwei Ye,[1,*] Yaroslav V. Kartashov,[2,3] Victor A. Vysloukh,[4] and Xianfeng Chen[1]

[1]*Key Laboratory for Laser Plasma (Ministry of Education), Collaborative Innovation Center of IFSA, Department of Physics and Astronomy, Shanghai Jiao Tong University, Shanghai 200240, China*
[2]*ICFO-Institut de Ciencies Fotoniques, The Barcelona Institute of Science and Technology, 08860 Castelldefels (Barcelona), Spain*
[3]*Institute of Spectroscopy, Russian Academy of Sciences, Troitsk, Moscow, 142190, Russian Federation*
[4]*Universidad de las Americas Puebla, Santa Catarina Martir, 72820, Puebla, Mexico*
*\*Corresponding author: fangweiye@sjtu.edu.cn*

**We show that truncated rotating square waveguide arrays support new types of *localized* modes that exist even in the linear case, in complete contrast to localized excitations in nonrotating arrays requiring nonlinearity for their existence and forming above the energy flow threshold. These new modes appear either around array center, since rotation leads to the emergence of the effective attractive potential with a minimum at the rotation axis, or in the array corners, in which case localization occurs due to competition between centrifugal force (in terms of quasi-particle analogy) and total internal reflection at the interface of the truncated array. The degree of localization of the central and corner modes mediated by rotation increases with rotation frequency. Stable rotating soliton families bifurcating from linear modes are analyzed in both focusing and defocusing media.**

***OCIS codes:*** *(190.0190) Nonlinear optics; (190.6135) Spatial solitons.*

http://dx.doi.org/10.1364/OL.99.099999

Generation of localized long-range excitations in the depth or at the surface of linear and nonlinear periodic materials is a problem of continuously renewed interest in photonics, since such states inherit unusual dispersion properties of Bloch eigenmodes and may be used for various practical applications, including information transfer, waveform and diffraction control, sensing, and surface characterization, to name just a few [1,2]. Especially interesting in this respect are surface states, whose properties depend not only on the microstructure of periodic material, but are determined also by the medium placed in contact with it.

Many different types of surface waves at the interfaces of periodic materials have been reported to the date. They include linear waves at the interfaces of photonic crystals with high refractive index contrast, forming at particular optical frequencies and having propagation constants in the forbidden gaps [3-5] by analogy with electronic Tamm or Shockley states [6,7]. Linear surface waves may form at the interfaces of specially designed shallow optical lattices [8,9], where optical analogs of electronic Shockley states were encountered [10]. Such modes appear also at the interfaces of materials with different topologies, including interfaces between uniform medium and truncated honeycomb lattices [11,12], where longitudinal modulations of the structure may lead to topological protection of surface states. Longitudinal bending of waveguides in the array may lead to linear near-surface dynamic localization observed in [13].

Surface waves at the interface of periodic structures may exist due to nonlinearity of the material, as observed in [14,15]. If mean refractive index in the uniform medium and in the lattice is different, surface waves usually appear above certain flow threshold – a characteristic feature that was discovered in seminal papers on surface waves at uniform interfaces [16,17] (see also reviews [18-20]). A variety of nonlinear surface states observed experimentally include gap [21-24] and different two-dimensional [25-27] surface solitons. The impact of surface geometry on such waves was analyzed in [28,29].

A different approach to formation of surface waves was suggested in [30], where it was shown that *rotation* allows localizing light at the interfaces of *radially symmetric* lattices even in the linear regime. This mechanism of surface wave formation, that is qualitatively different from previously discussed mechanisms, relies on competition between centrifugal energy transfer and total internal reflection at the array interface. The existence of rotating surface modes was confirmed experimentally in circularly symmetric optical cavities [31].

The aim of this Letter is to show that this mechanism can support linear corner (and central) modes in more complex structures – rotating square waveguide arrays –featuring only discrete, but not continuous rotation symmetry. We start from discussion of linear modes supported by the rotating array, and then switch to solitons bifurcating from them. We stress that previously light localization was studied only near the center of rotating periodic arrays [32-35]. To the best of our knowledge, rotating linear and nonlinear modes were never obtained explicitly as entities remaining invariable in the coordinate frame co-rotating with the array.

We consider the propagation of light beams along the $x$ axis in a medium with cubic nonlinearity, and the evolution of the dimensionless field amplitude $q$ is governed by the nonlinear Schrodinger equation:

$$i\frac{\partial q}{\partial \xi}=-\frac{1}{2}\left(\frac{\partial^2 q}{\partial \eta^2}+\frac{\partial^2 q}{\partial \zeta^2}\right)+\sigma q|q|^2 - R(\eta,\zeta,\xi)q, \qquad (1)$$

where the longitudinal $\xi$ and transvers $\eta$, $\zeta$ coordinates are normalized to the diffraction length and the input beam width, respectively; $\sigma = \mp 1$ corresponds to focusing/defocusing nonlinearity ($\sigma = 0$ in linear medium); the array $R(\eta,\zeta,\xi)$ is built from super-Gaussian waveguides $p\exp\{-[(\eta-\eta_k)^2+(\zeta-\zeta_m)^2]^2/w^4\}$ arranged into square structure with separation $d$ between neighboring sites, where $p$ is the array depth proportional to real refractive index variation. The array rotates as a whole around the origin at $(\eta,\zeta)=0$ with angular frequency $a$, i.e. waveguide center positions $(\eta_k,\zeta_m)$ are the harmonic functions of the distance $\xi$. We assume that the array is truncated and consists of 19×19 waveguides. Further we set the period of the array to $d = 2$, waveguide widths to $w = 0.5$, and array depth to $p = 8$. The waveguide width and depth are selected such that the waveguides are single-mode. Experimentally, such rotating waveguide arrays may be written in glass by focused femtosecond laser pulses [11]. Twisted multi-core optical fibers may serve as another candidate for experimental observation of states reported here [2]. The array considered here features $C_{4v}$ discrete rotation symmetry, however, similar results are anticipated in truncated arrays with other symmetry types, including honeycomb ones.

We are interested in *stationary modes* that rotate together with the array. In order to find them, we move to the rotating coordinate frame $\eta'=\eta\cos(\alpha\xi)+\zeta\sin(\alpha\xi)$, $\zeta'=\zeta\cos(\alpha\xi)-\eta\sin(\alpha\xi)$ in which waveguide array does not change and assume that modes have nontrivial phase profiles $q=[u(\eta',\zeta')+iv(\eta',\zeta')]\exp(ib\xi)$ (here $u$, $v$ are the real and imaginary parts of the field amplitude, and $b$ is the propagation constant. In the rotating coordinate system the equations for the field components take the form (for simplicity, we further omit primes in the coordinates):

$$\alpha(\eta\partial/\partial\zeta-\zeta\partial/\partial\eta)v+(1/2)\Delta_\perp u-\sigma u(u^2+v^2)+ Ru=bu \\ \alpha(\zeta\partial/\partial\eta-\eta\partial/\partial\zeta)u+(1/2)\Delta_\perp v-\sigma v(u^2+v^2)+ Rv=bv \qquad (2)$$

where $\Delta_\perp = \partial^2/\partial\eta^2+\partial^2/\partial\zeta^2$ is the transverse Laplacian accounting for diffraction. In the frames of the effective particle description, the first advective terms in Eqs. (2) account for centrifugal force acting on the beam [the analogy between Eq. (1) written in the rotating coordinate frame and two-dimensional nonrelativistic Schrödinger equation for a charged particle moving in the periodic potential and subjected to time-independent electric and magnetic fields is established in Ref. [35]], third and fourth terms stand for self-action and refraction in optically inhomogeneous medium. We characterize solutions of Eq. (2) with their energy flow $U = \iint |q|^2 d\eta d\zeta$, which is among the conserved quantities of Eq. (1).

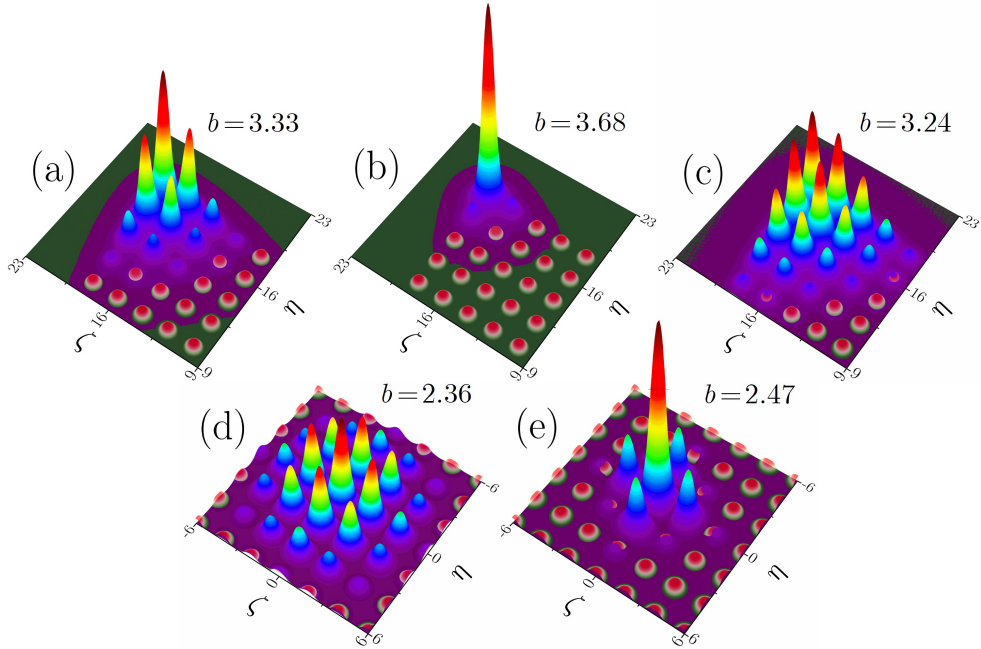

Fig. 1. (Color online) (a)-(c) Linear and nonlinear modes residing in the corner of rotating array at $a = 0.046$. (a) Linear mode with $b = 3.33$, (b) soliton in focusing medium with $b = 3.68$, (c) soliton in defocusing medium with $b = 3.24$. Linear modes residing in the center of rotating array at (d) $b = 2.36$, $a = 0.08$ and (e) $b = 2.47$, $a = 0.26$. Small red spots in all panels indicate array channels.

We first consider *linear modes* of Eq. (2) at $s = 0$ that can be obtained with standard eigenvalue solver. At $a = 0$, i.e. when waveguide array does not rotate, all linear modes are in fact delocalized. Because our array is finite and contains 19×19 waveguides, the transverse extent of all modes at $a = 0$ is determined by the size of the entire array. The central result of this Letter is that for $\alpha \neq 0$ two types of localized linear modes emerge. One of them is located in the array corner (more precisely, due to equiva-lence of all four corners of the array, they appear in each corner of the structure). An example of such a mode is shown in Fig. 1(a). The physical reason behind the existence of such modes is an interplay of centrifugal light energy transfer and total internal reflection at the array border. Indeed, the former energy transfer acts such that all off-center excitations are expelled toward the edge of the array. At the same time, light tends to be reflected back into the depth of array when it reaches array surface, since the mean refractive index in the array is higher than that in the surrounding free space. The mode depicted in Fig. 1(a) has highest propagation constant and is most confined, but other less confined modes may appear in the corner too. They have different phase distributions and correspond to lower $b$ values.

In addition to *corner* mode, localized linear states emerge in the *center* of the waveguide array. Examples of such states are shown in Figs. 1(d),(e). They appear due to *averaging* of the rotating square potential pointed out in [32] and leading to the formation of the minimum of the effective potential at the rotation axis. This effective potential becomes deeper with the increase of the rotation frequency. As a result, linear central modes corresponding to higher rotation frequency [Fig. 1(e), $a = 0.26$] are more localized than modes at low frequency [Fig. 1(d), $a = 0.08$].

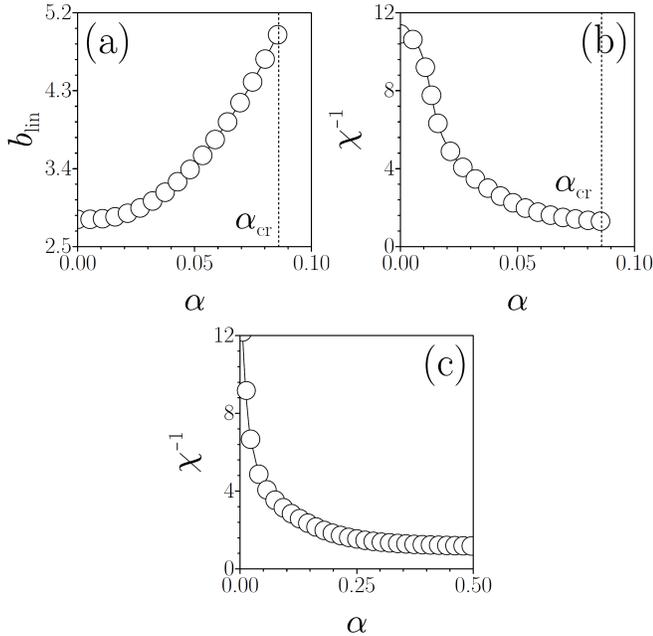

Fig. 2. (Color online) (a) Propagation constant of linear corner mode existing due to rotation and (b) its width versus $a$. Dashed lines indicate critical rotation frequency. (c) Width of linear central mode versus $a$.

The propagation constant of the most localized linear *corner* mode (termed here $b_{\text{lin}}$) coincides with the top of the first band at $a = 0$ and monotonically increases with the increase of the rotation frequency [Fig. 2(a)]. Growth of the rotation frequency is accompanied by the monotonic localization of corner mode (that might be interpreted as a consequence of increased centrifugal force pushing light toward the surface). The width of the mode $W = \chi^{-1}$ can be characterized using integral form-factor $\chi^2 = U^{-2} \iint |q|^2 d\eta d\zeta$ that provides accurate estimate even for complex field distributions like those depicted in Fig. 1. Fig. 2(b) shows that for sufficiently large $a$ values the corner mode shrinks nearly to a single waveguide excitation. When rotation frequency exceeds a critical value $\alpha_{\text{cr}} \approx 0.084$ marked with dashed line in Fig. 2(a), no localized modes residing in the array corners can be found, since the potential cannot compensate centrifugal light energy transfer at such frequencies. Instead of corner modes at $a > a_{\text{cr}}$ the eigenvalue solver returns modes located near the boundary of the integration window. We found that $a_{\text{cr}}$ monotonically grows with the increase of the array depth $p$ (thus, at $p = 11$ one gets $\alpha_{\text{cr}} \approx 0.11$) and decreases with the increase of its period $d$ (since larger $d$ leads to increase of the array size and growth of centrifugal forces).

*Central* modes may withstand much larger rotation frequencies than corner ones [see Fig. 2(c) showing transformation of the width of such modes with $a$]. However, at sufficiently large $a$ values exceeding $0.5$ the central modes become leaky and acquire small-amplitude background. The transition between localized and leaky central modes seems to be continuous and it is hard to introduce well-defined critical rotation frequency for them. Nevertheless, direct propagation shows that radiation is negligible for central mode with $a = 0.3$.

Next we address nonlinear modes supported by truncated rotating array at $\sigma \neq 0$. In our case, nonlinear modes bifurcate from linear ones upon increase of the peak amplitude (i.e. linear limit corresponds to $|q|, U \to 0$). They can be encountered both in focusing and in defocusing medium. Typical $U(b)$ dependencies are shown in Fig. 3(a) for different rotation frequencies, including $a = 0$ case. In the non-rotating array the localized corner modes exist only above certain energy flow threshold (curve 1). The manifestation of this threshold is still seen at small rotation frequencies, where dependence $U(b)$ may be nonmonotonic (curve 2). At sufficiently large $a > 0.034$ the energy flow of corner mode monotonically decreases with $b$ in defocusing medium and increases in focusing medium (curve 3). Increasing $b$ in focusing medium results in further mode contraction towards corner channel [compare Fig. 1(b) showing nonlinear corner mode with its linear counterpart from Fig. 1(a)]. Decreasing $b$ in defocusing medium leads to notable expansion of the corner mode [Fig. 1(c)]. In focusing medium for a fixed propagation constant $b$, the energy flow of soliton decreases with increase of rotation frequency [Fig. 3(b)] and at certain $a = a_{\text{cr}}$ value the energy flow vanishes, so that soliton transforms into linear mode. Corresponding critical rotation frequency can be defined from Fig. 2(a). Since solitons bifurcate from linear guided modes, their width remains finite in the bifurcation point and is determined by the width of corresponding linear mode [Fig. 2(b)]. Stability analysis performed by direct soliton propagation in the presence of random perturbations shows that at intermediate rotation frequencies $a < a_{\text{cr}}$ nonlinear corner modes are stable. Instabilities are possible at $\alpha \to 0$ only on the $U(b)$ branches with negative slopes. Notice that similarly to corner modes, solitons can bifurcate from linear modes located in the center of the array. Their properties are analogous to those of corner modes, so we do not consider them in details here.

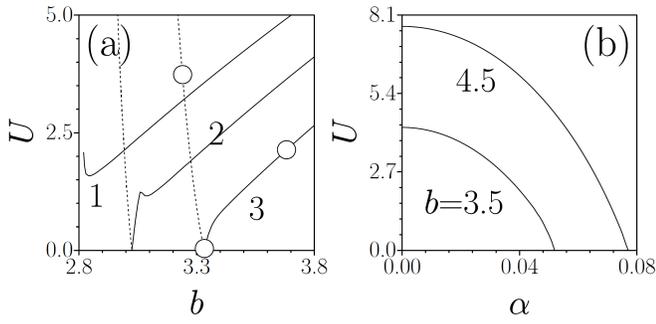

Fig. 3. (Color online) (a) Energy flow versus propagation constant for corner solitons in arrays with $a = 0$ (curve 1), $a = 0.032$ (curve 2), and $a = 0.046$ (curve 3). Solid lines correspond to focusing nonlinearity, dashed lines to defocusing one. Circles correspond to corner modes shown in Fig. 1. (b) Energy flow versus $a$ for different $b$ values in focusing medium.

Example of stationary rotation of exact linear corner mode obtained from Eq. (2) is depicted in Fig. 4(a), where light intensity distributions at different propagation distances are superimposed for clear visualization of the trajectory. Such mode indeed rotates steadily without radiation as long as $a < a_{cr}$. Since corner supports several linear modes with different phase and amplitude distributions, a single-site input $q|_{\xi=0} = a\exp[-(\eta+9d)^2/w^2 - (\zeta+9d)^2/w^2]$ with small amplitude $a = 1$ excites several such modes with corresponding weight coefficients, hence subsequent evolution is a result of intermodal beatings [see Fig. 4(b), $a = 0.01$]. Upon such beatings the light remains close to the corner that was excited and does not penetrate into array center. When focusing nonlinearity is considered and the amplitude $a$ is comparable with $1$ the rotating corner soliton is excited [see Fig. 4(c), $a = 1$]. Notice that almost no radiation penetrates into the depth of array.

Summarizing, we have shown that rotating waveguide arrays are capable of supporting stationary corner and central linear and nonlinear modes that rotate without detectable radiation in proper parameter region. Rotation eliminates energy flow threshold required for existence of surface solitons at the edge of periodic medium.

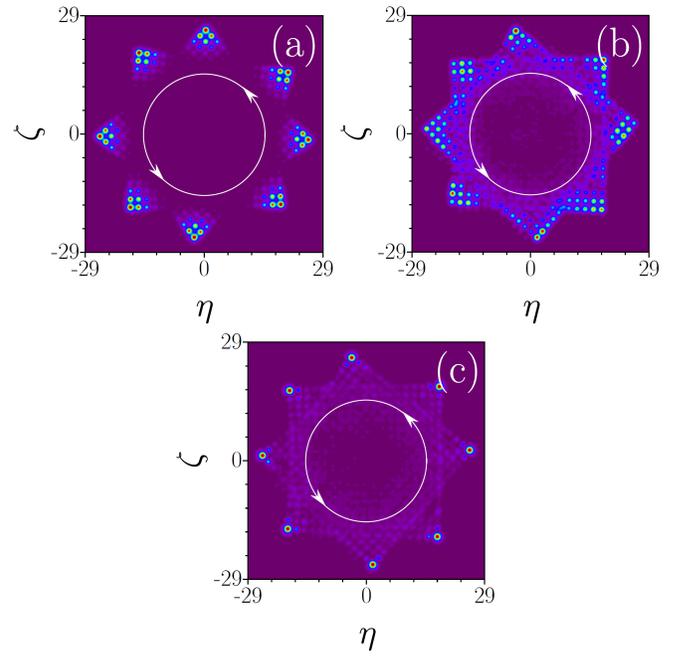

Fig. 4. (Color online) Snapshot images showing evolution of exact linear corner mode in rotating array (a), linear evolution of localized beam launched into corner channel (b), and excitation of corner soliton in focusing medium with localized beam launched into corner channel (c). In all cases only one corner is excited and $a = 0.046$. Snapshot images are taken at distances $x = \{0, 15, 35, 55, 70, 85, 105, 120\}$ in panel (a), and at distances $x = \{135, 155, 170, 190, 205, 225, 240, 255\}$ in (b) and (c). White circle with arrows indicate rotation direction.